\documentclass[aps,pra,showpacs,twocolumn,amsmath,amssymb]{revtex4-1}

\usepackage{graphicx}
\usepackage{dcolumn}
\usepackage{bm}
\usepackage{amsmath}
\usepackage{amssymb}
\usepackage{latexsym}
\usepackage{epsfig}
\usepackage{amsbsy}
\usepackage{array}
\usepackage{amssymb}
\usepackage{setspace}
\usepackage{bm}

\begin{document}

\title{Finite time cooling in dispersively and dissipatively coupled optomechanics}

\author{Tian Chen$^{1,2}$}

\author{Xiang-Bin Wang$^{1,2,3}$}\email{xbwang@mail.tsinghua.edu.cn}

\affiliation{
$^1$State Key Laboratory of Low
Dimensional Quantum Physics, Department of Physics, Tsinghua University, Beijing 100084,
People's Republic of China\\
$^2$Synergetic Innovation Center of Quantum Information and Quantum Physics, University of Science and Technology of China, Hefei, Anhui 230026, People's Republic of China\\
$^3$Jinan Institute of Quantum Technology, Shandong
Academy of Information and Communication Technology, Jinan 250101,
People's Republic of China
}
\begin{abstract}
The cooling performance of an optomechanical system comprising both dispersive and dissipative coupling is studied. We present a scheme to cool a mechanical resonator to its ground state in finite time by employing a chirped pulse. When the cavity damping strength increases, the phonon occupation of the resonator will decrease. Moreover, the cooling behaviors of this dispersively and dissipatively coupled system with different incident pulses, different system coupling strengths are explored. Our scheme is feasible to cool the resonator in a wide parameter region.
\end{abstract}


\maketitle

\section{introduction}
Cooling a mechanical resonator to its ground state plays a fundamentally important role in ultrahigh sensitive detection, observation of the transition between the quantum and classical system, and so on. Thus it has drawn much attention of researchers~\cite{Clerk_RMP, Aspelmeyer_Rev}. In many versatile designs, an auxiliary system is introduced to cool the mechanical resonator. The ground state cooling of a mechanical resonator can be realized by coupling the resonator with a driven cavity via radiation pressure~\cite{Law_PRA}, e.g. backaction cooling via a detuning cavity, cold-damping quantum feedback cooling \cite{Marquardt_PRL, Wilson-Rae_PRL, Wilson-Rae_NJP, Genes_PRA}, and other methods \cite{Tian_PRB, Wang_PRL11, Machnes_PRL, Eberly_book, Hioe_PRA, Liao_PRA, Gu_PRA, Liu_PRL, Liu_CPB}. By employing a chirped pulse, which relates the self induced transparency in the lossless medium, the resonator can be cooled in finite time \cite{Eberly_book, Hioe_PRA, Liao_PRA}. Also, cooling a mechanical resonator has been experimentally implemented \cite{Arcizet_nature, Gigan_nature, Schliesser_natphys08, Groblacher_natphys, Schliesser_natphys09, Park_natphys, Rivire_PRA, Teufel_nature, Verhagen_nature}. When combing the sideband cooling technique and cryogenic cooling, the mechanical resonator can be cooled beyond the nanoscale, and the position of the resonator has been measured \cite{Arcizet_nature, Gigan_nature, Schliesser_natphys08, Groblacher_natphys, Schliesser_natphys09, Park_natphys, Rivire_PRA}. Recently, a quantum coherent exchange between the mechanical object and micro/optical wave is achieved \cite{Teufel_nature, Verhagen_nature}, and the phonon occupation of the resonator is cooled to around one.

For an experimental realization of the optomechanical scheme, it is not enough to consider the optomechanical coupling by the radiation pressure only~\cite{Li_Natph}. The motion of the resonator affects the cavity damping significantly, hence it is worth of studying a system of which mechanical motion modulates both the intracavity photon number (dispersive coupling) and the cavity damping strength (dissipative coupling) \cite{Elste_PRL, Weiss_PRA, Weiss_NJP}. In recent experiments, a microwave optomechanical system is applied to cool a mechanical resonator~\cite{Regal_Natp, Teufel_PRL}. Such a system has advantages that it is operated well into the resolved-sideband limit and is compatible with precooling. As pointed out in Ref.~\cite{Elste_PRL}, besides the dispersive coupling, there also exists dissipative coupling for such a system. The experiment with a microdisk-waveguide optomechanical system has demonstrated that, the force from the dissipative coupling dominates the total force applied to the waveguide \cite{Li_PRL}. Due to these two different couplings, the quantum destructive interference emerges. The optimal detuning of the incident optical field frequency in stationary state cooling is dependent on the ratio of these two coupling strengths. As shown in \cite{Elste_PRL}, the cooling of the resonator does not require the good-cavity limit. Experimental proposals for the two different couplings have been put forward \cite{Xuereb_PRL, Yan_PRA}.

However, all of these works above focus on the steady-state cooling. Here, we discuss the cooling of such a dispersively and dissipatively coupled system in \textit{finite time}. We show that, by modulating the incident pulses, the phonon occupation of the resonator can be lowered to one. We study the cooling behavior of the mechanical resonator for the system consisting of the dispersive and dissipative coupling in finite time. When a chirped pulse is applied, better performance in cooling of the mechanical resonator appears with the increase of the cavity damping strength. This is different from the case there is only dispersive coupling, where a better cooling appears with a smaller cavity damping strength, as shown by Liao and Law~\cite{Liao_PRA}. Considering the tunability of the optomechanical dispersive and dissipative coupling strengths~\cite{Xuereb_PRL, Yan_PRA}, we show the different cooling performances of the resonator with a chirped pulse. Moreover, we modulate the strength of frequency sweep of the chirped pulse to explore the cooling effect. We find that the chirped pulse can be used to cool resonators efficiently in finite time in a large class of system parameter setting. We also study the steady-state phonon number of the resonator without chirping.

The structure of this paper is as follows, Sec.~II introduces the dispersively and dissipatively coupled optomechanics system we shall study. In Sec.~III, we explore the cooling performance with different cavity damping strengths, different ratios of the dispersive and dissipative coupling strengths, and different strength of frequency sweeping field. The paper is concluded by Sec.~IV.

\section{model and solution}
We consider an optomechanical system that consists of a mechanical resonator and a cavity mode. Taking both the dispersive coupling and the dissipation coupling into account, one has the following total Hamiltonian is~\cite{Elste_PRL},
\begin{equation}
\begin{split}
H=&\omega_c a^\dagger a+\omega_m b^\dagger b+H_\kappa+H_\gamma\\
&-(A\kappa a^\dagger a+i\sqrt{\frac{\kappa}{2\pi\rho}}\frac{B}{2}\sum_q(a^\dagger b_q-b_q^\dagger a))(b^\dagger +b).\label{H}
\end{split}
\end{equation}
Here, we have used the notations as following, $\omega_c$: the cavity frequency, $\omega_m$: the resonance frequency of the mechanical oscillator, $a$ $(a^\dagger)$: the annihilation (creation) operator of the cavity mode, $b$ $(b^\dagger)$: the annihilation (creation) operator of the mechanical resonator, $H_\kappa$ and $H_\gamma$: the damping of the cavity and mechanical resonator, respectively, $A$ ($B$): the dispersive (dissipation) coupling strength, $\kappa$: the cavity damping strength, $b_{q}$ $(b_{q}^{\dagger})$: the annihilation (creation) operator of the optical bath coupled to the cavity mode, $\rho$: the density of state for the optical bath. By employing the input-output formalism \cite{Walls_book}, we can obtain,
\begin{equation}
\sqrt{\frac{\kappa}{2\pi\rho}}\sum_q b_q=\sqrt{\kappa}a_{in}+\frac{\kappa}{2}a+\frac{\kappa}{2}\frac{B}{2}(b+b^\dagger )a,
\end{equation}
where $a_{in}$ is the input mode. Using $a=(\langle a\rangle+\delta a)e^{-i\omega_d t}$, $b=\langle b\rangle+\delta b$, and $a_{in}=(\langle a_{in}\rangle+\xi_{in})e^{-i\omega_d t}$, $\omega_d$ is the frequency of the driven field, $\xi_{in}$ is the noise induced by the optical bath. We shall use the following coupled differential equations of in frame rotating at the driven frequency~\cite{Weiss_NJP},
\begin{subequations}
\begin{align}
\dot{\delta a}=&(i\Delta+i2A\kappa \Re[\langle b(t)\rangle]-\frac{\kappa}{2}-\kappa B \Re[\langle b(t)\rangle])\delta a+(iA\kappa\langle a(t)\rangle\notag\\&-\frac{\kappa}{2}B\langle a(t)\rangle-i\Omega(t)\frac{B}{2})\delta b^\dagger +(iA\kappa\langle a(t)\rangle-\frac{\kappa}{2}B\langle a(t)\rangle\notag\\&-i\Omega(t)\frac{B}{2})\delta b-\sqrt{\kappa}(1+B \Re[\langle b(t)\rangle])\xi_{in},\label{del1}\\
\dot{\delta b}=&(-i\omega_m-\frac{\gamma}{2})\delta b+(iA\kappa\langle a(t)\rangle^*-i\frac{B}{2}\Omega(t)^*)\delta a\notag\\&+(iA\kappa\langle a(t)\rangle-i\frac{B}{2}\Omega(t))\delta a^\dagger -\frac{B}{2}\sqrt{\kappa}\langle a(t)\rangle^*\xi_{in}\notag\\&+\frac{B}{2}\sqrt{\kappa}\langle a(t)\rangle\xi_{in}^\dagger -\sqrt{\gamma}\eta.\label{del2}
\end{align}
\end{subequations}
Here, $\Delta=\omega_d-\omega_c$: the detuning of the cavity drive, $\delta a$ ($\delta b$): the fluctuation component of the cavity mode (mechanical mode), $\langle a\rangle$: the intra-cavity amplitude, $\Omega(t)=-i\sqrt{\kappa}\langle a_{in}\rangle$: the amplitude of the coherent laser drive, $\eta$: the noise influencing the mechanical resonator, $\gamma$: the mechanical intrinsic damping strength, $\Re[\langle b(t)\rangle]$: the real part of $\langle b(t)\rangle$. The time evolution equations for $\langle a\rangle$ and $\langle b\rangle$ are,
\begin{subequations}
\begin{align}
\langle\dot{a(t)}\rangle=&i\Delta\langle a(t)\rangle+i2A\kappa\Re[\langle b(t)\rangle]\langle a(t)\rangle-\kappa B\Re[\langle b(t)\rangle]\langle a(t)\rangle\notag\\&-i\Omega(t)-iB\Re[\langle b(t)\rangle]\Omega(t)-\frac{\kappa}{2}\langle a(t)\rangle,\label{del3}\\
\langle\dot{b(t)}\rangle=&-i\omega_m\langle b(t)\rangle+iA\kappa|\langle a(t)\rangle|^2-\frac{\gamma}{2}\langle b(t)\rangle\notag\\&-i\frac{B}{2}(\Omega(t)\langle a(t)\rangle^*+\Omega(t)^*\langle a(t)\rangle).\label{del4}
\end{align}
\end{subequations}
For simplicity, we will use some succinct expressions below,
\begin{subequations}
\begin{align}
F_1(t)=&i\Delta+i2A\kappa\Re[\langle b(t)\rangle]-\frac{\kappa}{2}-\kappa B\Re[\langle b(t)\rangle],\\
F_2(t)=&iA\kappa\langle a(t)\rangle-\frac{\kappa}{2}B\langle a(t)\rangle-i\frac{B}{2}\Omega(t),\\
F_3(t)=&\sqrt{\kappa}(1+B\Re[\langle b(t)\rangle]),\\
F_4(t)=&iA\kappa\langle a(t)\rangle^*-i\frac{B}{2}\Omega(t)^*.
\end{align}
\end{subequations}
We set $\vec{V}(t)=[\delta a,\delta b,\delta a^\dagger ,\delta b^\dagger ]^T$, the dynamics of $\vec{V}(t)$ is satisfied by,
\begin{equation}
\dot{\vec{V}}(t)=M(t)\vec{V}(t)+\vec{N}(t).
\end{equation}
with,
\begin{equation}
M(t)=\left(
\begin{array}{cccc}
F_1(t) & F_2(t) & 0 & F_2(t)\\
F_4(t) & -i\omega_m-\frac{\gamma}{2} & -F_4(t)^* & 0\\
0 & F_2(t)^* & F_1(t)^* & F_2(t)^*\\
-F_4(t) & 0 & F_4(t)^* & i\omega_m-\frac{\gamma}{2}
\end{array}
\right),
\end{equation}
and the noise related terms,
\begin{equation}
\vec{N}(t)=\left(
\begin{array}{c}
-F_3(t)\xi_{in}\\
-\frac{B}{2}\sqrt{\kappa}\langle a\rangle^*\xi_{in}+\frac{B}{2}\sqrt{\kappa}\langle a\rangle\xi_{in}^\dagger -\sqrt{\gamma}\eta\\
-F_3(t)^*\xi_{in}^{+}\\
-\frac{B}{2}\sqrt{\kappa}\langle a\rangle\xi_{in}^\dagger +\frac{B}{2}\sqrt{\kappa}\langle a\rangle^*\xi_{in}-\sqrt{\gamma}\eta^\dagger
\end{array}
\right).
\end{equation}
We introduce a covariance matrix notation $R_{l,l^{'}}(t)=\langle v_l(t)v_{l^{'}}(t^{'})\rangle$ $(l,l^{'}=1,2,3,4)$, $v_l$ is the $l$th element of $\vec{V}$. The dynamics of $R$ can be obtained as,
\begin{equation}
R(t)=G(t)R(0)G(t)^T+G(t)Z(t)G(t)^T,\label{R_mat}
\end{equation}
here, $Z(t)$ is,
\begin{equation}
Z(t)=\int_0^t\int_0^t G(\tau)^{-1}C(\tau,\tau^{'})[G(\tau^{'})^-1]^T d\tau d\tau^{'},
\end{equation}
with $G(t)$ being governed by $\dot{G}(t)=M(t)G(t)$, $G(0)$ is an identity matrix, $C_{l,l^{'}}(\tau,\tau^{'})=\langle N_{l}(\tau)N_{l^{'}}(\tau^{'})\rangle$ $(l,l^{'}=1,2,3,4)$, $N_{l}$ is the $l$th element of $\vec{N}$. Considering the Markovian bath, we can write $C(\tau,\tau^{'})$ in the form $C(\tau)\delta(\tau-\tau^{'})$, and the matrix expression of $C(\tau)$ is given below,
\begin{widetext}
\begin{equation}
C(\tau)=\left(
\begin{array}{cccc}
0 & -F_3(\tau)\frac{B}{2}\sqrt{\kappa}\langle a(\tau)\rangle & |F_3(\tau)|^2 & F_3(\tau)\frac{B}{2}\sqrt{\kappa}\langle a(\tau)\rangle\\
0 & -\frac{B^2}{4}\kappa|\langle a(\tau)\rangle|^2 & \frac{B}{2}\sqrt{\kappa}F_3(\tau)^*\langle a(\tau)\rangle^* & \frac{B^2}{4}\kappa|\langle a(\tau)\rangle|^2+\gamma(N_{th}+1)\\
0 & 0 & 0 & 0\\
0 & \frac{B^2}{4}\sqrt{\kappa}|\langle a(\tau)\rangle|^2+\gamma N_{th} & -\frac{B}{2}\sqrt{\kappa}F_3(\tau)^*\langle a(\tau)\rangle^* & -\frac{B^2}{4}\kappa|\langle a(\tau)\rangle|^2
\end{array}
\right).
\end{equation}
\end{widetext}
Here, $N_{th}$ is the distribution of the thermal bath surrounding the mechanical resonator. We assume that the initial optical bath is a zero-temperature bath, the initial condition $R(0)$ has three nonzero elements, $R_{13}(0)=1$, $R_{24}(0)=N_{th}+1$, and $R_{42}(0)=N_{th}$. Based on the definition of the covariance matrix $R_{l,l^{'}}(t)$, we can obtain the time-dependent mean displaced phonon number as $\langle\delta b^\dagger \delta b\rangle=R_{42}(t)$, and the mean displaced photo number $\langle\delta a^\dagger \delta a\rangle=R_{31}(t)$.

\textit{Chirped pulse form}. Here, we take the incident pulse as a chirped pulse form, which can lead to phenomenon of the self induced transparency in the lossless medium \cite{Eberly_book, Hioe_PRA}. This chirped pulse form has been applied to cool an optomechanical system consisting of dispersive coupling between the mechanical resonator and the cavity mode efficiently. The optomechanical linear coupling can be described by the Hamiltonian $[\langle a(t)^\dagger\rangle\delta a+\langle a(t)\rangle\delta a^\dagger](\delta b^\dagger+\delta b)$ \cite{Liao_PRA}. In essence, after applying the rotation wave approximation (RWA), this linear coupling Hamiltonian has the same form as the form of the optical-matter interaction, and the pulse can totally convert the population of two-level-system (TLS) \cite{Eberly_book, Hioe_PRA}. This is the reason why the chirped pulse has a good cooling performance. In the system we are studying, there are both dispersive coupling and dissipative coupling. We apply the chirped pulse to cool this complicated system. Based on the discussion above, we set, $A\kappa\langle a(t)\rangle-\Omega(t)\frac{B}{2}=\chi(t)\cdot e^{i\phi(t)}$, $\chi(t)=\chi_0\cdot\mathrm{sech}(\alpha(t-t_0))$, the frequency sweep $\dot{\phi}(t)=\beta\mathrm{tanh}(\alpha(t-t_0))$. The time-dependent phonon number can be obtained from the derivations above, with $\Omega(t)$ being substituted by such chirped pulse form.

\section{results}
\begin{figure}[htbp]
\begin{center}
\includegraphics[width=0.5\textwidth]{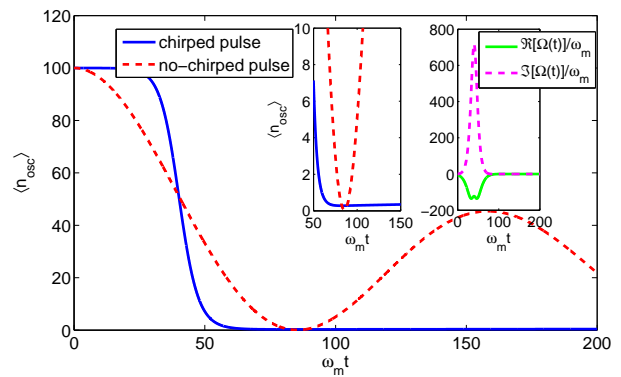}
\end{center}
\caption{\label{fig1} (Color online): The time evolution of the phonon number $\langle n_{osc}\rangle$=$\langle\delta b^\dagger \delta b\rangle$. Blue solid one denotes the chirped pulse form, that is $A\kappa\langle a(t)\rangle-\Omega(t)\frac{B}{2}=\chi(t)e^{i\phi(t)}$, red dashed one is the no-chirped form. Other parameters: $A=0$, $B=2*10^{-4}$, $\kappa/\omega_m=0.01$, $\gamma/\omega_m=10^{-5}$, $N_{th}=100$, $\langle a(0)\rangle=200$, $\Delta/\omega_m=-1$. For the chirped pulse form, $\chi(t)=\chi_0\cdot\mathrm{sech}(\alpha(t-t_0))$, $\dot{\phi}(t)=\beta\mathrm{tanh}(\alpha(t-t_0))$. We use $\alpha/\omega_m=0.14$, $\beta/\omega_m=0.04$, $\chi_0=\frac{1}{2}\sqrt{\alpha^2+\beta^2}$, $\omega_m t_0=40$.}
\end{figure}

We numerically calculate Eq.~(\ref{R_mat}), and obtain the residual phonon number of the mechanical resonator ($\langle b^\dagger  b\rangle$) with the time evolution. Two different schemes are provided for comparisons. One uses the chirped pulse (chirped pulse scheme), the other does not use the chirped pulse (no-chirped pulse scheme). We find that at time $\omega_mt=80$ or longer, the phonon number can nearly keep stable for the chirped pulse cooling scheme. While, for the no-chirped pulse scheme, the mechanical resonator is affected by the intrinsic mechanical damping and quantum backaction from the driven cavity. The phonon number changes more drastically, as shown in Fig.~\ref{fig1}. Although some smaller occupation can be achieved in the no-chirped pulse case at some time, the occupation will quickly raise to a high value. The figure inset shows the time evolution of $\Omega(t)$ for the chirped pulse scheme.

\begin{figure}[htbp]
\begin{center}
\includegraphics[width=0.5\textwidth]{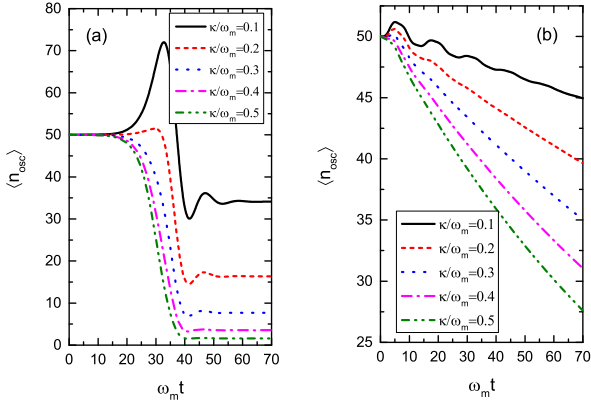}
\end{center}
\caption{\label{fig2} (Color online): The time evolution of the phonon number $\langle n_{osc}\rangle$=$\langle\delta b^\dagger \delta b\rangle$, with five different cavity damping $\kappa$. Left: the chirped pulse form, $A\kappa\langle a(t)\rangle-\Omega(t)\frac{B}{2}=\chi(t)e^{i\phi(t)}$, right: the no-chirped form. Other parameters: $A=0$, $B=2*10^{-4}$, $\gamma/\omega_m=10^{-6}$, $N_{th}=50$, $\langle a(0)\rangle=10^3$, $\Delta/\omega_m=0.5$, which is the optimal detuning. For the chirped pulse form, $\chi(t)=\chi_0\cdot\mathrm{sech}(\alpha(t-t_0))$, $\dot{\phi}(t)=\beta\mathrm{tanh}(\alpha(t-t_0))$. Parameters set, $\alpha/\omega_m=0.15$, $\beta/\omega_m=0.05$, $\chi_0=\frac{3}{2}\sqrt{\alpha^2+\beta^2}$, $\omega_m t_0=30$.}
\end{figure}

In Fig.~\ref{fig2}, we explore the cavity damping effect on the residual phonon number of the mechanical resonator. The chirped pulse scheme with five different cavity damping strengths is shown in Fig.~\ref{fig2}.~(a), when the cavity dampling $\kappa/\omega_m=0.5$, the mechanical resonator number ($\langle\delta b^\dagger \delta b\rangle$) achieves $1.6$ at time $\omega_mt=70$. In Fig.~\ref{fig2}.~(b), we apply a no-chirped pulse scheme. In our setting, the cavity detuning is $\Delta/\omega_m=0.5$. Again, we find that the phonon number decreases with the cavity damping strength. It is noted that the phonon number of the resonator is affected by two physical processes, one is the occupation transfer from the resonator to the zero-temperature cavity mode, the other is dissipations of the mechanical resonator and the cavity mode into their surrounded baths. When the cavity damping strength raises, occupation transfer from the resonator to the cavity mode becomes more possible. So in the larger cavity damping case, we can obtain a much smaller phonon occupation for the resonator in \textit{finite time}. Similar to Ref.~\cite{Liao_PRA}, the photon number of this dispersively and dissipatively coupled system increases from zero to a peak value, and then decrease to zero gradually, which is not shown here explicitly. Moreover, when comparing Fig.~\ref{fig1} and Fig.~\ref{fig2}, we find that the chirped pulse scheme is advantageous in cooling the mechanical resonator in short time.

\begin{figure}[htbp]
\begin{center}
\includegraphics[width=0.5\textwidth]{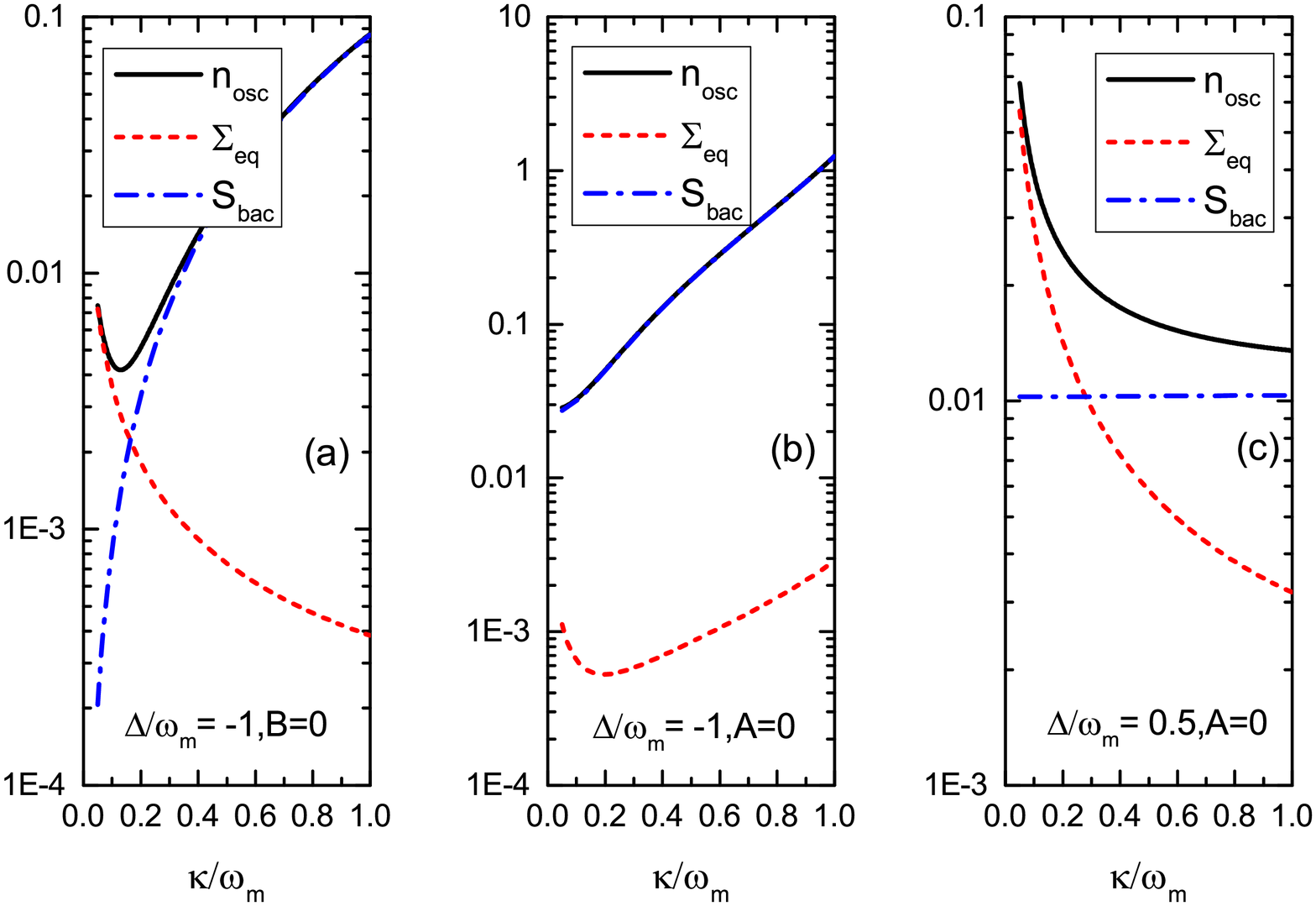}
\end{center}
\caption{\label{fig3} (Color online): The contributions to the steady-state phonon number $\langle n_{osc}\rangle$=$\langle\delta b^\dagger \delta b\rangle$ from two terms $\Sigma_{eq}$ (intrinsic oscillator damping), $S_{bac}$ (backaction from the cavity), with the change of cavity damping $\kappa/\omega_{m}$. (a): pure dispersion case, $A=2*10^{-4}$, $B=0$, $\Delta/\omega_m=-1$, (b): pure dissipation case, $A=0$, $B=2*10^{-4}$, $\Delta/\omega_m=-1$, (c): pure dissipation case, $A=0$, $B=2*10^{-4}$, $\Delta/\omega_m=0.5$. For all of three plots, $\gamma/\omega_m=10^{-6}$, $N_{th}=50$, intra-cavity amplitude at the steady state, $\langle a\rangle=10^3$.}
\end{figure}

In the no-chirped pulse scheme, the mechanical intrinsic damping ($\Sigma_{eq}$) and quantum backaction force from the driven cavity ($S_{bac}$) contribute to the final occupation as,
\begin{equation}
\begin{split}
\langle n_{osc}\rangle&=\int\frac{d\omega}{2\pi}S_{cc}(\omega)=\Sigma_{eq}+S_{bac}\\
&=\int\frac{d\omega}{2\pi}\frac{\gamma\sigma_{th}(\omega)}{|N(\omega)|^2}+\int\frac{d\omega}{2\pi}\frac{\kappa\sigma_{opt}(\omega)}{|N(\omega)|^2},\\
\end{split}
\end{equation}
where, $S_{cc}(\omega)$ is the mechanical spectrum, $\sigma_{th}(\omega)$ is a quantity related to the mechanical thermal bath and optomechanical self-energy, $\sigma_{opt}(\omega)$ combines the intracavity amplitude and quantum backaction force spectrum. $N(\omega)$ comprises the response function of the mechanical resonator and optomechanical self-energy. $\kappa$ and $\gamma$ are the optical damping strength and mechanical intrinsic damping strength, respectively. The detailed expression can be found in \textit{cf.} Eq.~(16) \cite{Weiss_NJP}. The numerical results of no-chirped pulse scheme is shown in Fig.~\ref{fig3}.

The pure dispersion case is shown in Fig.~\ref{fig3}.~(a), and the cavity detuning is $\Delta/\omega_m=-1$. Fig.~\ref{fig3}. (b) and (c) are the pure dissipation case with cavity detunings $\Delta/\omega_m=-1$ and $\Delta/\omega_m=0.5$, respectively. When the cavity damping rises, the steady-state phonon number due to the mechanical intrinsic damping will decrease, however, the contribution from backaction force leads to the fast increase of the final occupation. Considering these two opposite effects, when the cavity damping strength increases, the occupation of the resonator decreases first, then rises up quickly. The cavity damping is modulated by the resonator, when the cavity detuning is not optimal, the occupation is mainly determined by the quantum backaction force, see Fig.~\ref{fig3}.~(b). While, when the cavity detuning is optimal (Fig.~\ref{fig3}.~(c)), the contribution from the intrinsic mechanical damping decreases with the increase of the cavity damping, but the phonon occupation due to the quantum backaction force increases slowly. It is noted that for the case of pure dissipative coupling, the term $\sqrt{\frac{\kappa}{2\pi\rho}}\frac{B}{2}\sum_q(a^\dagger b_q-b_q^\dagger a)(b^\dagger +b)$ from Eq.~(\ref{H}) influences the cavity damping, and reduces the intracavity photon number. Based on these, compared with the pure dispersion case, the backaction effect will be weaken in the dissipative case. When combining the contributions from intrinsic mechanic damping and quantum backaction force, the residual phonon number of the resonator will decrease with the increase of the cavity damping. This result is consistent with that in Fig.~\ref{fig2}.

\begin{figure}[htbp]
\begin{center}
\includegraphics[width=0.5\textwidth]{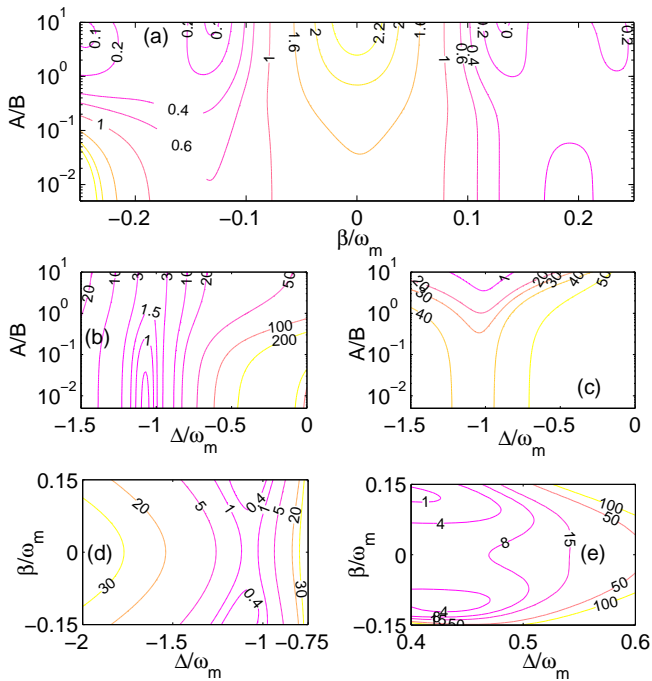}
\end{center}
\caption{\label{fig4} (Color online): The chirped pulse form, $\chi(t)=\chi_0\cdot\mathrm{sech}(\alpha(t-t_0))$, $\dot{\phi}(t)=\beta\mathrm{tanh}(\alpha(t-t_0))$, $\chi_0=\frac{3}{2}\sqrt{\alpha^2+\beta^2}$. (a): the effect of the optomechanical coupling strength ratio $(A/B)$ and the strength of frequency sweeping field $\beta/\omega_m$. $\alpha/\omega_m=0.15$, $\Delta/\omega_m=-1$. (b): the effect of the optomechanical coupling strength ratio $(A/B)$ and cavity detuning $\Delta/\omega_m$. $\alpha/\omega_m=0.15$, $\beta/\omega_m=0.05$. (c) is the no-chirped pulse scheme, to compares with (b). (d) and (e): the strength of frequency sweeping field $\beta/\omega_m$ and cavity detuning $\Delta/\omega_m$ to the phonon number $\langle n_{osc}\rangle$=$\langle\delta b^\dagger \delta b\rangle$. $A=0$, $\alpha/\omega_m=0.15$. For all of the plots, the time is set $\omega_m t=70$. $B=2*10^{-4}$, $\gamma/\omega_m=10^{-6}$, $\kappa/\omega_m=0.3$, $N_{th}=50$, $\langle a(0)\rangle=200$, $\omega_mt_0=30$.}
\end{figure}

Now, we discuss the occupation of the resonator in the chirped pulse scheme with different settings below. The evolution time is $\omega_mt=70\geq2t_0$. Firstly, the effect of different strengths of frequency sweeping field and optomechanical coupling strength ratios (dispersive coupling strength/dissipative coupling strength) is shown in Fig.~\ref{fig4}.~(a). The phonon occupation of the resonator distributes symmetrically around $\beta/\omega_m=0$. In the region of $0.08\leq|\beta/\omega_m|\leq0.18$, the residual occupation is lower than one. This result is insensitive (fault tolerant) to the strength of sweeping field. It is well known that the dispersive coupling and dissipative coupling strength can be modulated in the designed scheme \cite{Xuereb_PRL, Yan_PRA}. So we can achieve the ground state cooling of the mechanical resonator in a wide parameter region.

Secondly, we explore the effects of the cavity detunings. The residual phonon number of the mechanical resonator with the chirped pulse scheme and no-chirped pulse scheme are compared in Fig.~\ref{fig4}.~(b) and (c). For the chirped pulse case (b), we can reduce the phonon occupation of the resonator to one in the vicinity of $\Delta/\omega_m=-1$, with the coupling strength ratio $A/B\leq0.2$. While, the resonator can be cooled to around one phonon with the coupling strength ratio $A/B\geq5$ in the no-chirped pulse scheme (c). When we choose the ratio $A/B\geq5$ and $\Delta/\omega_m\simeq-1$ in the chirped pulse case, the obtained phonon occupation of the resonator is smaller than three. Although it might not likely to provide the cooling performance as effective as that of the no-chirped pulse scheme in the region $A/B\geq5$, the chirped pulse scheme can cool the resonator efficiently in a much wider region, $5*10^{-3}\leq A/B\leq10$.


Finally, in the pure dissipation case, the influence of the frequency sweep and cavity detuning to the phonon occupation is shown in Fig.~\ref{fig4}.~(d) and (e). The better cooling effect appears when the frequency sweep value $|\beta/\omega_m|$ is relatively large. In the region around $\Delta/\omega_m=-1$ and $\Delta/\omega_m=0.4$, we can reduce the residual phonon occupation of the resonator below one when the strength of sweeping field is chosen $\beta/\omega_m=0.125$.

\section{conclusion}
In summary, we have extensively studied the cooling of a mechanical resonator for the system consisting of \textit{both} dispersive coupling and dissipative coupling in \textit{finite time}. By employing a chirped pulse scheme, the mechanical resonator is cooled, and the phonon occupation of the resonator keeps stable in relatively long time. We find that, in our optomechanical system, the better cooling performance appears with the increase of the cavity damping strength. Our proposal with a chirped pulse can be used to cool resonators efficiently in finite time in a large range of system parameter settings.

\section*{Acknowledgement}
 We acknowledge the financial support in part by the 10000-Plan of Shandong province, and the National High-Tech Program of China grant No. 2011AA010800 and 2011AA010803, NSFC grant No. 11174177 and 60725416.
{}

\begin{thebibliography}{}

\bibitem{Clerk_RMP} A. A. Clerk, M. H. Devoret, S. M. Girvin, F. Marquardt, and R. J. Schoelkopf, Rev. Mod. Phys. \textbf{82}, 1155 (2010).

\bibitem{Aspelmeyer_Rev} M. Aspelmeyer, T. J. Kippenberg, and F. Marquardt, arXiv:1303.0733v1.

\bibitem{Law_PRA} C. K. Law, Phys. Rev. A \textbf{51}, 2537 (1995).

\bibitem{Marquardt_PRL} F. Marquardt, J. P. Chen, A. A. Clerk, and S. M. Girvin, Phys. Rev. Lett. \textbf{99}, 093902 (2007).

\bibitem{Wilson-Rae_PRL} I. Wilson-Rae, N. Nooshi, W. Zwerger, and T. J. Kippenberg, Phys. Rev. Lett. \textbf{99}, 093901 (2007).

\bibitem{Wilson-Rae_NJP} I. Wilson-Rae, N. Nooshi, J. Dobrindt, T. J. Kippenberg, and W Zwerger, New. J. Phys. \textbf{10}, 095007 (2008).

\bibitem{Genes_PRA} C. Genes, D. Vitali, P. Tombesi, S. Gigan, and M. Aspelmeyer, Phys. Rev. A \textbf{77}, 033804 (2008).

\bibitem{Tian_PRB} L. Tian, Phys. Rev. B \textbf{84}, 035417 (2011).

\bibitem{Wang_PRL11} X. Wang, S. Vinjanampathy, F. W. Strauch, and K. Jacobs, Phys. Rev. Lett. \textbf{107}, 177204 (2011).

\bibitem{Machnes_PRL} S. Machnes, J. Cerrillo, M. Aspelmeyer, W. Wieczorek, M. B. Plenio, and A. Retzker, Phys. Rev. Lett. \textbf{108}, 153601 (2012).

\bibitem{Eberly_book} L. Allen and J. H. Eberly, \textit{Optical Resonance and Two-Level Atoms} (Dover, New York, 1987).

\bibitem{Hioe_PRA} F. T. Hioe, Phys. Rev. A \textbf{30}, 2100 (1984).

\bibitem{Liao_PRA} J-Q. Liao and C. K. Law, Phys. Rev. A \textbf{84}, 053838 (2011).

\bibitem{Gu_PRA} W-J. Gu and G-X. Li, Phys. Rev. A \textbf{87}, 025804 (2013).

\bibitem{Liu_PRL} Y-C. Liu, Y-F. Xiao, X. Luan, and C. W. Wong, Phys. Rev. Lett. \textbf{110}, 153606 (2013).

\bibitem{Liu_CPB} Y-C. Liu, Y. W. Hu, C. W. Wong, and Y-F. Xiao, Chin. Phys. B  \textbf{22}, 114213 (2013).

\bibitem{Arcizet_nature} O. Arcizet, P.-F. Cohadon, T. Briant, M. Pinard, and A. Heidmann, Nature \textbf{444}, 71 (2006).

\bibitem{Gigan_nature} S. Gigan, H. R. B\"{o}hm, M. Paternostro, F. Blaser, G. Langer, J. B. Hertzberg, K. C. Schwab, D. B\"{a}uerle, M. Aspelmeyer, and A. Zeilinger, Nature \textbf{444}, 67 (2006).

\bibitem{Schliesser_natphys08} A. Schliesser, R. Rivi\`{e}re, G. Anetsberger, O. Arcizet, and T. J. Kippenberg, Nat. Phys. \textbf{4}, 415 (2008).

\bibitem{Groblacher_natphys} S. Gr\"{o}blacher, J. B. Hertzberg, M. R. Vanner, G. D. Cole, S. Gigan, K. C. Schwab, and M. Aspelmeyer, Nat. Phys. \textbf{5}, 485 (2009).

\bibitem{Schliesser_natphys09} A. Schliesser, O. Arcizet, R. Rivi\`{e}re, G. Anetsberger, and T. J. Kippenberg, Nat. Phys. \textbf{5}, 509 (2009).

\bibitem{Park_natphys} Y-S. Park and H. Wang, Nat. Phys. \textbf{5}, 489 (2009).

\bibitem{Rivire_PRA} R. Rivi\`{e}re, S. Del\'{e}glise, S. Weis, E. Gavartin, O. Arcizet, A. Schliesser, and T. J. Kippenberg, Phys. Rev. A \textbf{83}, 063835 (2011).

\bibitem{Teufel_nature} J. D. Teufel, T. Donner, D. Li, J. W. Harlow, M. S. Allman, K. Cicak, A. J. Sirois, J. D. Whittaker, K. W. Lehnert, R. W. Simmonds, Nature \textbf{475}, 359 (2011).

\bibitem{Verhagen_nature} E. Verhagen, S. Del\'{e}glise, S. Weis, A. Schliesser, and T. J. Kippenberg, Nature \textbf{482}, 63 (2012).

\bibitem{Li_Natph} M. Li, W. H. P. Pernice, and H. X. Tang, Nat. Photon. \textbf{3}, 464 (2009).

\bibitem{Elste_PRL} F. Elste, S. M. Girvin, and A. A. Clerk, Phys. Rev. Lett. \textbf{102}, 207209 (2009); F. Elste, S. M. Girvin, and A. A. Clerk, Phys. Rev. Lett. \textbf{103}, 149902 (2009).

\bibitem{Weiss_PRA} T. Weiss and A. Nunnenkamp, Phys. Rev. A \textbf{88}, 023850 (2013).

\bibitem{Weiss_NJP} T. Weiss, C. Bruder, and A. Nunnenkamp, New. J. Phys. \textbf{15}, 045017 (2013).

\bibitem{Regal_Natp} C. A. Regal, J. D. Teufel, and K. W. Lehnert, Nat. Phys. \textbf{4}, 555 (2008).

\bibitem{Teufel_PRL} J. D. Teufel, J. W. Harlow, C. A. Regal, and K. W. Lehnert, Phys. Rev. Lett. \textbf{101}, 197203 (2008).

\bibitem{Li_PRL} M. Li, W. H. P. Pernice, and H. X. Tang, Phys. Rev. Lett. \textbf{103}, 223901 (2009).

\bibitem{Xuereb_PRL} A. Xuereb, R. Schnabel, and K. Hammerer, Phys. Rev. Lett. \textbf{107}, 213604 (2011).

\bibitem{Yan_PRA} M-Y. Yan, H-K. Li, Y-C. Liu, W-L. Jin, and Y-F. Xiao, Phys. Rev. A \textbf{88}, 023802 (2013).

\bibitem{Walls_book} D. F. Walls and G. J. Milburn, \textit{Quantum Optics} (Springer, Berlin, 1994).

\end{thebibliography}
\end{document}